\documentclass[11pt]{article}
\usepackage[table,xcdraw]{xcolor}
\usepackage{multirow}
\usepackage{graphicx}
\usepackage{tikz}
\usetikzlibrary{decorations.pathmorphing}
\usetikzlibrary{shapes.misc}
\usetikzlibrary{arrows}

\usepackage{floatrow}
\usepackage{multicol} 
\usepackage{color}
\usepackage{latexsym}
\definecolor{rosso}{cmyk}{0,1,1,0.4}

\definecolor{rossos}{cmyk}{0,1,1,0.55}
\definecolor{rossoc}{cmyk}{0,0.5,1,0.2}
\definecolor{blu}{cmyk}{1,1,0,0.3}
\definecolor{blus}{cmyk}{1,1,0,0.6}
\definecolor{blucc}{cmyk}{1,0.4,0.2,0}
\definecolor{viola}{cmyk}{0,1,0,0.6}
\definecolor{viola2}{cmyk}{0,1,0.2,0.6}
\definecolor{verde}{cmyk}{0.92,0,0.59,0.25}
\definecolor{verdec}{cmyk}{0.92,0,0.59,0.15}
\definecolor{verdes}{cmyk}{0.92,0,0.59,0.4}
\font\tenrsfs=rsfs10 at 12pt
\font\sevenrsfs=rsfs7
\font\fiversfs=rsfs5
\newfam\rsfsfam 
\textfont\rsfsfam=\tenrsfs
\scriptfont\rsfsfam=\sevenrsfs
\scriptscriptfont\rsfsfam=\fiversfs
\def\mathscr#1{{\fam\rsfsfam\relax#1}}

\def\circa#1{\,\raise.3ex\hbox{$#1$\kern-.75em\lower1ex\hbox{$\sim$}}\,}

\usepackage{amsmath,latexsym,amssymb,color,hyperref,graphicx}

\newcommand{\be}{\begin{equation}}
\newcommand{\ee}{\end{equation}}
\newcommand{\bea}{\begin{eqnarray}}
\newcommand{\ena}{\end{eqnarray}}
\newcommand{\no}{\noindent}

\renewcommand\k{\ensuremath{\kappa}}

\newcommand{\de}{\partial}
\newcommand{\ha}{\frac{1}{2}}

\newcommand{\ba}{\begin{eqnarray}}
\newcommand{\ea}{\end{eqnarray}}
\newcommand{\plm}{M_{pl}}

\newcommand{\LL}{\mathcal{L}}

\newcommand{\II}{\mathcal{I}}

\makeatletter
\def\ps@mine{%
    \def\@oddfoot{\hfil\thepage\hfil}\let\@evenfoot\@oddfoot
    \let\@oddhead\@evenhead%
    \let\@mkboth\@gobbletwo
    \let\sectionmark\@gobble
    \let\subsectionmark\@gobble
    }
\renewcommand\section{\@startsection {section}{1}{\z@}%
                                   {-3.5ex \@plus -1ex \@minus -.2ex}%
                                   {2ex \@plus.2ex}%
                                   {\normalfont\large\sffamily\bfseries}}
\renewcommand\subsection{\@startsection {subsection}{1}{\z@}%
                                   {-3.5ex \@plus -1ex \@minus -.2ex}%
                                   {2ex \@plus.2ex}%
                                   {\normalfont\sffamily\bfseries}}
                                   
                                   \renewcommand\subsubsection{\@startsection {subsubsection}{1}{\z@}%
                                   {-3.5ex \@plus -1ex \@minus -.2ex}%
                                   {2ex \@plus.2ex}%
                                   {\normalfont\sffamily\bfseries}}
\makeatother

\numberwithin{equation}{section}

\tikzset{
    photon/.style={decorate, decoration={snake}, draw=black} 
}
\tikzset{cross/.style={cross out, draw=black, minimum size=2*(#1-\pgflinewidth), inner sep=0pt, outer sep=0pt},
cross/.default={1pt}}

\usepackage{subcaption}
\begin{document}

\thispagestyle{empty}
\vspace*{-2.5cm}
\begin{minipage}{.45\linewidth}

\end{minipage}
\vspace{2.5cm}

\begin{center}
  {\huge\sffamily\bfseries Quantum Corrections to the Stochastic
    Gravitational Wave Background}
\end{center}
 
 \vspace{0.5cm}
 
 \begin{center} 
 {\sffamily\bfseries \large  Denis Comelli}$^{a}$,
{\sffamily\bfseries \large  Maicol Di Giambattista}$^{b,c}$, 
 {\sffamily\bfseries \large  Luigi Pilo}$^{b,c}$,
  {\sffamily\bfseries \large Rocco Rollo}$^{d}$\\[2ex]
$^a$ INFN, Sezione di Ferrara, I-44122 Ferrara, Italy \\\vspace{0.3cm}
$^b$ Dipartimento di Scienze Fisiche e Chimiche, Universit\`a degli Studi dell'Aquila,  I-67100 L'Aquila, Italy\\\vspace{0.3cm}
$^c$ INFN, Laboratori Nazionali del Gran Sasso, I-67100 Assergi, Italy\\\vspace{0.3cm}
$^d$ Centro Nazionale INFN di Studi Avanzati GGI, Largo Enrico Fermi 2,  I-50125 Firenze, Italy\\\vspace{0.3cm}
 {\tt comelli@fe.infn.it}, {\tt maicol.digiambattista@aquila.infn.it}, {\tt luigi.pilo@aquila.infn.it}, {\tt rocco.rollo@gssi.it},
\end{center}

\vspace{0.7cm}

 \begin{center}
 {\small \today}
 \end{center}

\vspace{0.7cm}

\abstract{We study 1-loop corrections to the primordial stochastic background of
  gravitational waves produced during inflation. While in
  single-clock, at the leading order in slow-roll, quantum corrections
  keep the amplitude scale-free this is not the case when the pattern of
  symmetry breaking is different. In particular, when spatial
  diffeomorphisms are also broken during inflation as for 
  solid inflation, a log-running in the external momentum is
  generated. We relate the appearance of a log-running to the
  spontaneous breaking of dilatation invariance. The running could be
  instrumental to distinguish single-clock from alternative
  models of inflation in future high sensitivity CMB polarisation and GWs experiments.   
}
\clearpage

\section{Introduction}
In the last couple of decades cosmology has become
sharper and sharper and the $\Lambda$CDM model is tested with great accuracy~\cite{Planck:2018vyg}. Within the inflationary
paradigm, the primordial perturbations have a quantum origin and
become ``classical'' only after horizon exit. Given that, it is useful to
test our ability to compute and predict
genuine quantum corrections to cosmological correlation functions. Such
a program was pioneered by Weinberg~\cite{Weinberg:2005vy} by using
the IN-IN formalism. Loop effects are notoriously subtle to compute in de
Sitter and quasi-de Sitter spacetime due to gauge and infrared
effects; see for instance~\cite{Seery:2010kh} for a review and a some
early reference. In particular, it was found~\cite{Adshead:2008gk,Adshead:2009cb} that in single
field inflation, the scalar and tensor 2-point functions get a sizeable correction of order
$\log(k)$ where $k$ is the comoving external momentum at the leading
order in slow-roll. Such a result was confuted
in~\cite{Senatore:2009cf,Senatore:2012nq,Pimentel:2012tw}, arguing
that after a proper implementation of the  regularisation procedure
actually  the correction $\log(k)$ to scalar power spectrum is turn
into a $\log(H)$ one and thus no
``running" is present. Here we reconsider the matter, focusing on the
tensor power spectrum   exploring also 
scenarios beyond single field inflation where the symmetry breaking
pattern is different. As a matter of fact there is a deep connection
between  the residual symmetry group of the inflationary background
and the presence of a $\log(k)$ running induced by quantum
corrections. During inflation the spacetime is close to a de Sitter
spacetime that has $SO(4,1)$ as isometry group. The underlying
conformal symmetry was used to derive~\cite{Creminelli_2012,Hinterbichler:2012nm,Hinterbichler:2013dpa,Hui:2018cag} the consistency relation in
single-field inflation~\cite{Maldacena:2002vr,Creminelli_2004,Cheung:2007sv}, to study systematically non-Gaussianity in
the tensor sector~\cite{Maldacena:2011nz} and scalar
sector~\cite{Creminelli:2011mw} for spectator fields. The idea is
that though in general dilation invariance is broken by vacuum
expectation value (VEV) of the inflaton, it remains unbroken in combination with a
suitable internal symmetry of the inflaton sector. This is the case in
single-clock inflation but when the symmetry breaking pattern is
different the running may appear. Such a running, besides its
theoretical interest, could be instrumental to improve our chances for
a direct detection of the primordial stochastic background of
gravitational waves. It should also be stressed that resurgence of
$\log(k)$ running also casts some doubts on the physical meaning of
the result at very small $k$ where perturbativity is at stake.

\section{Single Field Inflation}
\label{sf}
Consider single clock inflation based on a real scalar
field $\Phi$ minimally coupled to gravity described by the action
\be
S=\plm^2 \, \int d^4x \sqrt{-g} \left[ \frac{R}{2} +\, P(X, \Phi)
\right] \, ,
\qquad X= - \ha g^{\mu \nu} \de_\mu \Phi \de_\nu \Phi \,  .
\label{single}
\ee
In the spatially flat gauge, perturbations can be taken as
\be
\begin{split}
& ds^2 = \left(N_iN^i-N^2 \right) dt^2+ N_i \, dx^i dt + h_{ij} \, dx^i
dx^j \, ;\\
&h_{ij} =a^2 \left(e^{\gamma}\right)_{ij} \, , \qquad \qquad 
\Phi = \phi(t) + \pi(x) \, ;
\end{split}
\label{sfg}
\ee
where $\gamma_{ij}$ are transverse and traceless tensor perturbations
describing gravitational waves. The background metric is approximately
close to de Sitter (dS) spacetime for which~\footnote{We use
  conformal time $t$.} 
\be
a_{dS} = -\frac{1}{H \,  t} \, , \qquad  t\in ( -\infty, \, 0] \, .
\ee
Here $H$ is the Hubble rate in physical time (i.e. the time independent one in a pure dS background), which is related to the time dependent one expressed in conformal time $\mathcal{H}$ by\footnote{The prime denotes derivative with respect to the conformal time $t$.}
\be
\mathcal{H}=a \, H= a'/a
\ee
When $\phi' \neq 0$, a departure from a pure dS background is
is controlled by  the slow-roll parameter 
\be
\epsilon=\frac{
  \left(\mathcal{H}^2-\mathcal{H}'\right)}{\mathcal{H}^2} =\frac{P_X \, \phi'^2}{2 \, {\cal H}^2} \, ,
\ee
while the background field equations gives
\be
a^2 P+P_X  \, \phi'{}^2-3 \mathcal{H}^2 =0 \, .
\ee
Thus the kinetic part of the action for the inflaton is subdominant
with $\phi'$ small and\footnote{With a bar we denote the background
  values of a quantity.} $\bar{X}=\phi'{}^2/(2 a^2) $ almost constant in a quasi dS
regime. Indeed, the background equations are satisfied up to corrections of
order $\epsilon^2$ by taking
\be
\phi = \alpha \,   \log(a) \, , \qquad \qquad \qquad 
a= (- H \, t)^{(-1- \epsilon)} \, .
\ee
For instance taking $P(X,\Phi)=X^n-V(\Phi)$, with $n>0$, one can check
that $\alpha\sim \epsilon^{1/(2n)}$.
The background dS spacetime has 10 Killing vectors associated to the
generators of $SO(4,1)$; however,  the
inflaton classical configuration typically breaks spontaneously
$SO(4,1)$ down to spatial rotations and translation that are linearly realised. The spontaneous
broken dS symmetry is behind  the so called consistency relations~\cite{Maldacena:2002vr}
among correlation functions in the soft limit as shown in~\cite{Creminelli_2004,
  Cheung:2007sv,Creminelli:2012ed,Hinterbichler:2012nm,Hinterbichler:2013dpa}. Dilatations
and special conformal transformations of $SO(4,1)$ are non-linearly
realised and the underlying symmetry does not show up as invariance of
correlation functions as for rotations. Sometimes even dilations can be
linearly realised. Suppose now that $P$ has shift symmetry
\be
\Phi \to \Phi+c \, .
\label{ssym}
\ee
Such a symmetry is consistent with slow-roll $P$, according with 
the derivatives of $P$ with respect to $\Phi$ have to be small.
The background value of $ \Phi$ is not
invariant under dilatation; indeed, neglecting small slow-roll
corrections, we have that
\be
\phi \to \phi -  \alpha \, \log \lambda \, .
\label{transf}
\ee
However the effect of dilatation (\ref{transf}) on $\phi$ can be compensated by 
 a shift transformation of $\Phi$ by taking $c= \epsilon \, \alpha \, H^{-1} \log \lambda$. The combined action of a dilatation followed by a shift symmetry leaves
the background invariant. Thus, under a dilation we
 have for $\pi$
 \be
 \begin{split}
& \Phi'(x')=\Phi(x) \Rightarrow \phi(t)- \frac{\alpha}{H} \log(a)
+\pi'(x') = \phi(t) +\pi(x) \\
& \Rightarrow \pi'(x') =\pi(x) +\frac{\alpha}{H} \log(a) \, ;
\end{split}
 \ee
 while in  the ``diagonal'' combination of a dilatation followed by a
 shift symmetry (dshift), the nonlinear part of the trasformation is absent
 \be
 \pi'(x') =\pi(x)
 \ee
 and the dshift transformation is linearly realised.
 The same is true for curvature perturbation
of the 3D hypersurface orthogonal to $\de_\mu \Phi$ which is given by
the gauge invariant quantity $\cal{R} $ that for the metric in the
flat gauge (\ref{sfg}) is given by 
\be
\cal{R} = \frac{ {\cal H}}{\phi'}  \pi .
\ee
The quantity ${\cal R}$ is almost scale free constant on superhorizon
scales and it provides the seed for primordial perturbation. The quadratic Lagrangian for the field fluctuations $\cal{R}$ reads
in Fourier space
\be
L_2 =a^2 \, \plm^2 \, \epsilon\left[
  \frac{1}{ c_s^2}  \, \mathcal{R}'^2-k^2 \mathcal{R}^2\right]\, ,
\ee
with 
\be
c_s^2 =\frac{a^2 P_X}{a^2 P_X+P_{XX} \, \phi'{}^2} \, .
\label{csdef}
\ee
The cubic interaction with gravity is described by
\be
L_{TSS} = \plm^2 \, a^2 \, \epsilon \, \gamma_{ij} \, \de_i \mathcal{R} \,
\de_j \mathcal{R} \, .
\label{TSS}
\ee
The Lagrangian $L_2+L_{TSS}+L_{gw}$, where  $L_{gw}$ describes the
free dynamics of gravitons, is invariant under a dshift and such a
symmetry is linearly realised up to small slow-roll corrections, with
both $\mathcal{R}$ and $\gamma_{ij}$ being dimensionless and with zero weight. 

The bottom line is that the tensor 2-point function is constrained to be
dilation invariant and has to be of the following form (see appendix \ref{dil})
\be
\langle \gamma_{ij}(t, \, \pmb{k}_1) \,  \gamma^{ij}(t, \, \pmb{k}_2)\rangle
=  ( 2 \pi)^3 \, \delta^{(3)}
\left(\pmb{k_1}+\pmb{k_2} \right)  \,\frac{ F(k_1 \, a_{\text{dS}}^{-1})}{k_1^3} \, , \qquad
|\pmb{k_1}|=k_1 \, ,
\label{ex2point}
\ee
where $F$ is an arbitrary function. Notice that (\ref{ex2point}) is
exact up to slow-roll correction and represents 
the power spectrum of stochastic gravitational waves
produced during inflation, that is one of the key prediction of
inflationary models.  The form (\ref{ex2point}) is incompatible with the 
presence of a $\log(k/\mu)$ dependence induced by loop corrections. Indeed, on
superhorizon scales $F$ becomes a constant.  
The actual
1-loop computation given in appendix (\ref{1loop}) confirms that for
$t \to 0$, no running is present and one finds
\be
P^{(\gamma)}= \frac{H^2}{4 \, \pi^2 \, \plm^2} \left[1+
  \frac{H^2}{\plm^2} \frac{\left(41+5 \, c_s^4-118 \, 
    c_s^2\right)}{240 \,  c_s^7}
\log \left(\frac{H}{\mu} \right) \right]  \, ,
\label{res1loop}
\ee
where $\mu$ is a physical scale introduced by dimensional
regularisation. Similar results were obtained
in~\cite{Bartolo:2010bu,delRio:2018vrj}. In the above expression we have kept only non-analytic
contributions, the divergent part are subtracted by suitable
counter-terms.
The curvature scale acts as  a natural infrared cutoff that prevents
the correction to become too large for small $k$. 
Notice that the correction is
local and analytical only on superhorizon
scales and thus there is no genuine local counter-term able to cancel it at
all scales. Finally, as a technical note, as pointed out
in~\cite{Senatore:2009cf}, for the cancelation it is
important to correct the modes wavefunctions when using dimensional
regularisation in de Sitter, in contrast with what happens in Minkowski spacetime. 

\section{Beyond Single-Field: Fluids and Solids}
\label{solids}
Things are different for what concerns dilation symmetry when we go beyond
single-field inflation. An interesting laboratory is 
inflation driven by a generic self-gravitating
medium~\cite{Celoria:2021cxq,Celoria:2019oiu,Celoria:2020diz} 
originally introduced as infrared modification of gravity~\cite{ussgf,Celoria:2017hfd} to
explain dark energy~\cite{Celoria:2017idi,
  Celoria:2017bbh}.  In particular, let us consider solid
inflation~\cite{Endlich:2012pz} based on
three scalar fields $\varphi^a$, 
$a=1\, ,\,2\, ,\, 3$  with  a shift symmetry~\cite{Dubovsky:2005xd,Celoria:2017bbh} 
\be 
\varphi^a \rightarrow \varphi^a +
c^a\, ,
\label{shift}
\ee
and  $SO(3)$ internal symmetry
\be
\varphi^a \to {\varphi'}^a= {\cal
  R}^a_{ \ b} \;\varphi^b \,,  \qquad a,b= 1,2,3, \qquad \qquad {\cal R}^t \;{\cal R} =1
\, .
\label{inrot}
\ee
Among the spacetime scalars shift symmetric operators with a single derivative of $\varphi^a$
\be
 B^{ab} = g^{\mu \nu} \de_\mu \varphi^a \, \de_\nu \varphi^b \;\;
 \qquad a,b=1,2,3  \, , 
\ee
one can extract 3  operators invariant under internal $SO(3)$
rotations (\ref{inrot}) 
\be
b=\sqrt{\text{Det}\left[\textbf{B}\right]}\, ,\qquad
\tau_X= \text{Tr}\left[\textbf{B}\right]\,,\qquad \tau_Y=
\frac{\text{Tr} \left[\textbf{B}^2\right]}{\tau_X{}^2},\qquad \tau_Z=
\frac{\text{Tr} \left[\textbf{B}^3\right]}{\tau_X{}^3} \, .
\label{ope}
\ee
Thus, we arrive at the action
\be 
S=\frac{\plm^2}{2} \, \int d^4x\, \sqrt{-g}\, R+\plm^2 \, \int d^4x \,
\sqrt{-g}\, U(b,\tau_Y,\,\tau_Z)\, .
\label{gact}
\ee
Notice that the case $U(b)$ describes an adiabatic
fluid\footnote{Actually, for a fluid the slow-roll regime does not exist.}~\cite{Ballesteros:2016kdx,Celoria:2017bbh}. 
The scalar fields have 
the VEV
\be
\bar \varphi^i = x^i \, .
\label{vac}
\ee
The existence of a spatially homogeneous background is allowed by the
presence of global symmetries of the scalar field action. 
Consider a special multi-field  model of inflation based on 
the ``vacuum'' configuration (\ref{vac}), which has a residual  global ``diagonal'' $ISO(3)$ symmetry. Indeed, a global spatial rotation ${\cal R}^i_{ \ j} x^j$ can be
absorbed by a corresponding inverse internal transformation of
$\varphi^a$ and the same is true for a global translation $x^i \to x^i +c^i$ thanks to
the shift symmetry  (\ref{shift}).

Let us focus on the scalar part by setting $\varphi^i = \de_i \pi_L $,
notice that $\pi_L$ has dimension $-2$.
By looking at the quadratic Lagrangian for the fluctuations one gets in
the flat gauge
\be
L_2 =\frac{1}{2} \,a^2 \,\plm^2\,{\cal H}^2\, \epsilon  \Big [
4 \, (\partial_{i}\pi_L')^2-
   4\,c_L^2 (\Delta\pi _L)^2 
   \Big ]+
\frac{ a^2\, M_{\text{pl}}^2}{4}\,\left[
     (\gamma_{ij}' \gamma_{ij}'-M_2\,a^2\,\gamma_{ij} \gamma_{ij} +
     \partial_{\k}\gamma_{ij} \de_k\gamma_{ij} \right] \, ,
 \label{slowact}
 \ee
 where $\Delta=\partial_i^2$ and $c_L^2$ is an effective sound speed
 defined by $c_L^2=-1+4 \, c_2^2/3$; the parameter $c_2$ is expressed in
 terms of the second derivatives of $P$ and controls the presence
 anisotropic stress in the energy momentum tensor which characterises
 a solid; one could also verify that it is nothing else that the sound speed of the transverse phonons. The curvature perturbation is related
 to $\pi_L$ in the flat gauge by
 \be
 \zeta = \frac{k^2}{3} \, \pi_L \, .
 \ee
Notice that the form of (\ref{slowact}) is very similar to the case of
a single field inflation with a notable exception: by dimensional
reasons, there is an overall ${\cal H}^2$ due to the fact that $\pi_L$
has dimension $-2$. Such a change is due to the different breaking
pattern: while in single field 4-dimensional diffeomorphisms are
spontaneously broken down to 3-dimensional spatial diffeomorphisms,
for solids time diffeomorphisms are unbroken while the spatial
ones are broken. While the Lagrangian  $L_2$ for solid inflation
together with the TSS interaction part is still invariant under
dilation by assigning to $\pi_L$ weight $-2$, the argument used for
$P(X,\Phi)$ does not work anymore. The nature of the breaking is such that shift symmetry
is unable to compensate the effect of a dilation. The change of the background
configuration (\ref{vac}) by a dilatation is
\be
\bar \varphi^i \to \lambda \, x^i
\ee
which is $x$-dependent and cannot be removed by (\ref{shift}). Another way to see this is the following argument: we can try to impose some internal symmetry in order to make the background configuration invariant under a dilation symmetry; what we get is that a necessary condition is that the symmetry required, consistent with the background is
\be
\pi_L\to\pi_L+\textbf{c}\cdot\textbf{x}
\ee
that is nothing else but the galilean shift also discussed
in~\cite{Creminelli_2012} that  is not well defined in a curved
background; therefore we expect the 2-point function be not invariant under dilation, since it probes scales comparable with $H$.
In addition, because of the
presence of three scalar fields the arguments typical of single field
according with the presence of the inflaton is equivalent to a universal
change of the time coordinate~\cite{Cheung:2007st} does not apply to a
self-gravitating media of the form (\ref{gact}). On a more practical side, while mode for the field
$\pi$ in the interaction picture in the single-field case is  a Bessel function of
order $3/2$ in the case of solid the mode of $\pi_L$ is a Bessel
function of order $5/2$. Indeed, the fact that both
graviton modes and $\pi$ have modes proportional to Bessel function of
order $3/2$ is instrumental in the cancellation of the log
running. A direct computation gives
\be
P^{(\gamma)} = \frac{H^2}{4 \, \pi^2 \, \plm^2} \left[1+
  \frac{H^2}{\plm^2} \, \mathcal{F}(c_L) \log \left(\frac{k}{\mu}\right)\right]  \, 
  \label{res1loopsolid}
\ee
which indeed features a log running in the external momentum $k$. We
stress again that we are working at the leading order in slow-roll and the
correction has pure quantum origin. This last point has to be
carefully considered when higher-order corrections in perturbation
theory are studied by analysing the classical equations of motion in a
quasi de Sitter space, this is sometimes referred as gravitational
wave (GW)  backreaction
and the results in the literature disagree even in single field
models~\cite{Biagetti:2013kwa,Biagetti:2014asa,Fujita:2014oba}.
This topic and the role of quantum effects in a dS space during the
inflation-radiation transition will be the subject of future work.\\
For the complete form of the TSS interacting Hamiltonian and the $\mathcal{F}(c_L)$ function see Appendix \ref{loopsol}.
Finally, the fluid limit of the solid action (\ref{gact}) is rather interesting
as a consistency check. In a such limit only the operator $b$ is
present in  (\ref{gact}) and the internal symmetry is enhanced to
internal volume preserving
diffeomorphisms~\cite{Dubovsky:2005xd,ussgf,Ballesteros:2016kdx,Celoria:2017bbh}
and out of the three TSS interactions
terms~\cite{Endlich:2012pz,Celoria:2021cxq}, only a term of the form
\be
L_{TSS} = \plm^2 \, a^{4}_{dS}  \, \tilde A \, \gamma_{ij}  \,
  \de_i \zeta' \, \, \de_j  \zeta'  
  \, 
\label{gfluid}
\ee
survives, with $\zeta'\propto \de^2 \mathcal{R}/\mathcal{H}$.\\
As a matter of fact, the combination of modes (Hankel functions of order
5/2) and time derivatives of the vertex is such that it is completely
equivalent to the vertex in (\ref{TSS}) with $\nu=3/2$ and thus the
cancellation mechanism still works as it should be\footnote{The equivalence comes from the identity $\frac{d}{dx}\left[x^\nu \, H^{(1)}_\nu(x)\right]=x^\nu \, H^{(1)}_{\nu-1}(x)$.}. The reason behind
such non-trivial relation is that a perfect irrotational fluid described
in terms of three scalar field by the  Lagrangian $U(b)$ is related to
the description in terms of a single scalar by the Lagrangian $P(X)$
given in section \ref{sf}.

\section{Discussion and Conclusions}
The fact that in single field inflation there is no running in the
external momentum makes the quantum correction rather uneventful in the
sense that the amplitude of the power spectrum stays basically
scale-free at the leading order in slow-roll. Scale dependence is
reintroduced at 1-loop only by looking at the tiny effect of 
different modes exiting the horizon at slightly different times, when
slow-roll corrections to the scale factor $a$ are taken into
account (see~\cite{delRio:2018vrj}). Future CMB polarisation experiments can
hardly measure the simple dependence on the renormalisation scale.
An enhancement by a factor $N$ in the amplitude
can be obtained by considering $N$ scalar spectator fields and a
non-trivial sound speed.  Indeed the amplitude (\ref{res1loop}) scales with the sound speed as $c_s^{-7}$ and an even
stronger in the case of a supersolid~\cite{Celoria:2020diz}. Of course, the value of $c_s$ cannot be pushed toward too small values without jeopardising the validity of the derivative expansion in the effective theory of single field inflation and boosting the level of non-Gaussianity on the verge of conflicting with existing observations~\cite{Planck:2019kim}. 

When one moves away from the symmetry breaking pattern of single field
inflation the perspective is different. A genuine running with the
external momentum $k$ is indeed  present even at the leading order in
slow-roll and actually without any $\epsilon$ suppression. The running
is in principle significative enough to be used to distinguish the
breaking pattern (\ref{vac})  from  the one of single-field
inflation. On the down side, the result (\ref{res1loopsolid}) is
problematic in the limit $k \to 0$ and thus not infrared safe. Of course our analysis of 1-loop effect is not exhaustive,
further future investigations will establish whether the running is
one of many quirks of de Sitter spacetime or a clean physical prediction.

Things are different for future GW experiments. Even if suppressed by
slowroll parameters, the completely different log-scale dependence may
give a signature beyond the standard and suppressed contribution to
the stochastic GW background. In this case, we suggest that the
presence of a non trivial phonon sound speed (present in single field models with
a non-canonical kinetic term) could play an essential role in turning
the {\it red tilted} behaviour (see~\cite{Bartolo:2016ami,
  Guzzetti:2016mkm, Boyle_2008, Smith_2019}) in a {\it blue tilted}
one. A related point is the observation that the coefficient of the log in (\ref{res1loop}), a sort of beta function, is a monotonic function of $c_s$ that changes sign becoming negative for $c_s \gtrsim 0.6$. However, to find the complete form of this beta function, an analysis at the second order in slowroll expansion is mandatory during the dimensional regularisation procedure as argued in~\cite{delRio:2018vrj}. This will be the object of a future work.

 \vskip 1cm
 \no
 {\bf Acknowledgements}
 \\
 We thank Ema Dimastrogiovanni, Matteo Fasiello and Lucas Pinos for
 useful discussions and for sharing  the  draft of a
 forthcoming paper with some overlap with the present work and with
 similar results. 

\appendix

\section{Dilatation invariance}
\label{dil}
Take a generic field that  transforms as
\be
\Phi(x) \to \Phi'(x')= {\cal F}(\Phi)(x) \, ;
\ee
then the $n$-point function
\be
G_n(x_1, \cdots, x_n) =\langle \Phi(x_1) \cdots \Phi(x_n) \rangle
\ee
is such that
\be
G_n(x_1', \cdots, x_n') = \langle {\cal F}(\Phi)(x_1) \cdots  {\cal
  F}(\Phi)(x_n) \rangle \,  .
\ee
In particular for a field that is a scalar  (namely a dimensionless
scalar field) under the symmetry
\be
\Phi'(x')=\Phi(x)
\ee
and the correlation functions are invariant
\be
G_n(x_1, \cdots, x_n) =G_n(x_1', \cdots, x_n')  \, .
\label{inv}
\ee
Let us discuss the properties of a dimensionless scalar under dilatation. We are often
interested in the spatial Fourier transform of correlators; for
instance
\be
G_2(t_1,\pmb{x}_1; t_2,\pmb{x}_2) = \int \frac{d^3k_1}{\left( 2 \pi
  \right)^3} \frac{d^3k_2}{\left( 2 \pi \right)^3} \,
e^{ i \pmb{k_1} \cdot \pmb{x_1} +i \pmb{k_2}\cdot \pmb{x_2} } \,
\tilde{G}(t_1, \pmb{k_1}; t_2, \pmb{k_2}) \, .
\ee
In Fourier space (\ref{inv}) reads
\be
\tilde{G}(t_1, \pmb{k_1}; t_2, \pmb{k_2}) = \lambda^{-6} \,
\tilde{G}(\lambda \, t_1, \lambda \, \pmb{k_1}; \lambda \, t_2, \lambda
\, \pmb{k_2}) \, .
\label{Finv}
\ee
Let us focus on equal time correlators: $t_1=t_2=t$ and by
using spatial translation and rotational invariance of the dS metric one cast a
generic 2-point functions as
\be
\tilde{G}(t, \pmb{k_1}; t, \pmb{k_2}) \equiv \tilde{G}(t,
\pmb{k_1},  \pmb{k_2}) = ( 2 \pi)^3 \, \delta^{(3)}
\left(\pmb{k_1}+\pmb{k_2} \right)  \, F(k_1,t) \, , \qquad
|\pmb{k_1}|=k_1 \, .
\ee
By imposing (\ref{Finv}) to the 2-point function we have that
\be
\lambda^{-3} F(\lambda^{-1} \, k_1, \lambda \, t) = F(k_1,t)  \
;
\ee
which gives (\ref{ex2point})
\be
\tilde{G}(t,
\pmb{k_1},  \pmb{k_2}) = ( 2 \pi)^3 \, \delta^{(3)}
\left(\pmb{k_1}+\pmb{k_2} \right)  \, \frac{F(k_1 a_{dS}^{-1})}{k_1^3} \, .
\label{g2p}
\ee
\section{The In-In Formalism}
  \label{in-in}
Quantum correlators during inflation are typically computed by using the in-in
formalism expanding perturbatively the time evolution operator. A generic hermitian operator $W$ in
the Heisenberg picture is written as perturbative expansion in the
interaction picture 
\be
W(t) = {U(t)}^\dagger \, W_I(t) \, U(t) = W_I(t) +\sum_{n=1}^{\infty} W_I^{(n)}(t)  \, , \qquad \qquad U(t) = T \,
\exp \left[-i \int_{t_0}^{t}  dt' \, H_I(t') \right] \, ;
\ee
where $W_I(t)$ and  $H_I$ are the operator $W$ and  the relevant
interaction Hamiltonian are both in the
interaction picture. For the 1-loop corrections we will need the second order
term of the expansion that reads
\be
\begin{split}
W^{(2)}(t) &=
\int_{t_0^\ast}^{t} dt_2  \int_{t_0}^{t}
    dt_1 \, H_I(t_2) \, W_I(t) \, H_I(t_1) -\int_{t_0^\ast}^{t} dt_2 \, \int_{t_0^\ast}^{t_2}
    dt_1 \,  H_I(t_1) \, H_I(t_2) \, W_I(t) \\
    &- \int_{t_0}^{t} dt_2  \,
    \int_{t_0}^{t_2} dt_1 \,  W_I(t)  \,  H_I(t_2) \,  H_I(t_1) \, .
    \end{split}
    \, \label{in2}
 \ee
In dS in conformal time $t_0 \to - \infty$. Besides UV divergencies
due to virtual particles, in dS when the limit  $t_0 \to - \infty$ is
taken additional troubles appear and a regularisation procedure is
needed. An option is to take~\cite{Adshead:2008gk,Adshead:2009cb} $t_0=\bar t (1+ i \, \epsilon)$  with
$\bar t \to - \infty$,  to select the Bunch-Davies vacuum~\footnote{Of
  course the limit $\epsilon \to 0$ has to be taken afterwards.}. Thus
\be
\begin{split}
&\langle W(t) \rangle =
 2 \, \text{Re} \left[ \langle
  A(t) \rangle \right] + \langle B(t) \rangle \, ; \\
&
B(t)= \int_{t_0^\ast}^{t} dt_2  \int_{t_0}^{t}
    dt_1 \, H_I(t_2) \, W_I(t) \, H_I(t_1)  \, ,  \qquad A(t)= - \int_{t_0}^{t} dt_2  \,
    \int_{t_0}^{t_2} dt_1 \,  W_I(t)  \,  H_I(t_2) \,  H_I(t_1)  \, .
  \end{split}
  \label{comp}
\ee
It should be stressed that the above regularisation, due to the complex nature of $t_0$, forbids to
write down $W^{(n)}$ as a multiple commutator a la Weinberg~\cite{Weinberg:2005vy}. In order to do that one should
consider an adiabatic turn-off of
the interaction, see for instance~\cite{Baumgart:2020oby}
\be
 H_I(t) \to   \tilde H_I(t) =H_I(t) \, e^{\epsilon t}   \, \qquad
 \epsilon \in \mathbb{R} \, ,
 \label{rreg}
 \ee
to keep $U$ unitary.
 Thus, one can write
 \be
   W_I^{(2)}(t) = i^2 \, \int_{t_0}^{t} dt_2  \int_{t_0}^{t_2}
dt_1 \left[\tilde H_I(t_1), \, \left[\tilde H_I(t_2), \, W_I(t)
  \right] \right]  \, .
\label{wei}
\ee
While the form (\ref{comp}) leads to simpler time integrals, the
Weinberg form is more compact and allows to read off more easily the
late time behavior.

\section{Single Field Inflation}
\label{1loop}
To discuss the main features of the quantum corrections to the tensor
power spectrum consider single clock inflation based on a real scalar
field $\Phi$ described by action (\ref{single}). In the interaction picture we have  the following expressions for the fields
\be
\begin{split}
&\mathcal{R}(t,\pmb{x}) = \int \frac{d^3k}{\left( 2 \pi \right)^3}
\, e^{i\,\pmb{k} \cdot \pmb{x}} \left[a(\pmb{k}) \, f_k(t) + a^\dagger(-
    \pmb{k}) \,    f_k(t)^\ast \right] \equiv \int \frac{d^3k}{\left( 2 \pi \right)^3} \, e^{i
    \,\pmb{k} \cdot \pmb{x}} \mathcal{R}_{\pmb{k}}(t) \, ; \\
  &\gamma _{ij} (t,\pmb{x})= \sum_{s=1}^2 \int \frac{d^3k}{\left( 2 \pi \right)^3}
\, e^{i\,\pmb{k} \cdot \pmb{x}} \left[b(\pmb{k})_s \, \epsilon_{ij}^s h_k(t) + b^\dagger(-
    \pmb{k})_s \,  \, \epsilon_{ij}^s h_k(t)   \right] \equiv \int \frac{d^3k}{\left( 2 \pi \right)^3} \, e^{i
    \,\pmb{k} \cdot \pmb{x}} \gamma_{ij \, \pmb{k}}(t) \, .
\end{split}
\ee
In $3+\delta$ spatial dimensions, the wavefunctions for the scalar and tensor modes are given by
\be
f_{k}(t) =\frac{A_f}{k^{3/2}} \ (-H t)^{\delta/2} \ (-c_s k t)^{3/2} \ H_{(3+\delta)/2}^{(1)}(-c_s k t)
\label{Rmode}
\ee
\be
h_{k}(t) =\frac{A_{\gamma}}{k^{3/2}} \ (-H t)^{\delta/2} \ (-k t)^{3/2} \ H_{(3+\delta)/2}^{(1)}(-k t)
\label{hmode}
\ee
By expanding at first order in $\delta$ we get
\be
f_{k}(t) =\sqrt{\frac{2}{\pi}}\frac{A_f}{k^{3/2}} \ e^{-ic_skt}(-i+c_skt)+\delta\frac{A_f}{k^{3/2}\sqrt{2\pi}} \ e^{-ic_skt}\left[(-i+c_skt)\log(-Ht)+u(c_s,k,t)\right]
\label{scalmod}
\ee
\be
h_{k}(t) =\sqrt{\frac{2}{\pi}}\frac{A_{\gamma}}{k^{3/2}} \ e^{-ikt}(-i+kt)+\delta\frac{A_{\gamma}}{k^{3/2}\sqrt{2\pi}} \ e^{-ikt}\left[(-i+kt)\log(-Ht)+u(c_s=1,k,t)\right]
\label{tenmod}
\ee
where
\be
u(c_s,k,t)=2 e^{2 i c_s k t} (c_s k t+i) (\text{Ei}(-2 i c_s k t)-i \pi )-i (4+\pi  (c_s k t-i))
\ee
with $\text{Ei}(z)$ exponential integral function. The actual 1-loop
computation leading to (\ref{res1loop}) is described in appendix \ref{nolog}.

\section{Solid Inflation: 1-Loop}
\label{loopsol}

The relevant cubic TSS vertex for solid inflation is in Fourier space (see~\cite{Celoria:2021cxq} for more details)
\be
\mathcal{L}=\epsilon \, \plm^2 \, H^2 \, a^4 \, \varepsilon_{ij}(\textbf{k})\left[V_1
  \, \hat{k}'^i \ \hat{k}'^j+V_2 \ (\hat{k}'\cdot\hat{k}'') \
  \hat{k}'^i \ \hat{k}''^j\right] \, h_k \, \zeta_{k'} \, \zeta_{k''} \, ,
\label{sollag}
\ee
where we have denoted a unit vector with an hat. The coefficients $V_1$ and $V_2$ can be written in terms of the solid Lagrangian $U(b,\tau_Y,\tau_Z)$ and its derivatives, see~\cite{Celoria:2021cxq}.\\
Performing the momentum integrals by using dimensional regularisation
we get  (\ref{res1loopsolid}), where
\be
\begin{split}
\mathcal{F}(c_L)=&-\frac{1}{6531840 \, c_L^7}\left[ (83181 -4938\, c_L^2+25 \, c_L^4-40320 \gamma_E) \, V_1^2 \right. \\
 \ & +2 \, (83181-13968 \, c_L^2+355 \, c_L^4-40320 \, \gamma_E) \, V_1 \, V_2\\
 \ & \left. +(83181-22998 \, c_L^2+1105 \, c_L^4-40320 \, \gamma_E) \, V_2^2\right] \, .
\end{split}
\ee
Actually, by expanding the action (\ref{gact}) at the cubic
order~\cite{Celoria:2021cxq}, there exists another TSS  of the form
\be
\LL=\epsilon \, \plm^2 \, H^2 \, a^4 \, V_3 \, \varepsilon_{ij}(\textbf{k}) \, \frac{\hat{k}'^i \, \hat{k}''^j}{k' \, k''} \, h_k \, \zeta'_{k'} \, \zeta'_{k''} \, .
\label{extra}
\ee
The above vertex can be brought into the same form of (\ref{TSS}),
with the scalar and tensor modes have the same index for the Hankel
function $3/2$ as argued in section \ref{solids} and no logarithmic
running in the external momentum is produced.\\
However, one may ask if a mixed term coming from (\ref{sollag}) and
(\ref{extra}) could compete with (\ref{res1loopsolid}); after a
lengthy  computation one can see that the mixed terms proportional to
$V_1 V_3$ and $V_2 V_3$ are suppressed by extra powers of $c_L$, and
therefore are negligible for reasonable values of the (non trivial) effective sound speed $c_L$.

\section{Dimensional regularisation: beyond single field Inflation}
\label{nolog}

The structure of the logarithmic running coming from the loop
calculation can be inferred  without the need of evaluating explicitly
the momentum integrals by extending the argument given
in~\cite{Senatore:2009cf} for the case of standard single field
inflation with some mathematical assumptions. For TSS interactions, the contribution we need
to compute is of the form
\be
\begin{split}
\label{master_1}
 \left.\left\langle \hat{W}(t) \right\rangle\right|_{1-L} = &- 2\, \text{Re} \left[\int_{-\tilde\infty}^{t} dt_2\,\int_{-\tilde\infty}^{t_2} dt_1 \left\langle H_I^{(3)} (t_1)\,H_I^{(3)} (t_2)\, \hat{W}(t) \right\rangle\right] \\
 & +\int_{-\tilde\infty}^{t} dt_2\,\int_{-\tilde\infty^*}^{t} dt_1 \left\langle H_I^{(3)} (t_1)\, \hat{W}(t)\,H_I^{(3)} (t_2)\right\rangle.
\end{split}
\ee 
where $H_I$ is the interaction Hamiltonian. A regularisation procedure
is necessary in order to perform consistently the internal momentum
integral, which is divergent. By using dimensional regularisation, we
have to move to $3+\delta$ spatial dimensions and being in a curved
spacetime background also the form of the wavefunctions is modified;  the index of the Hankel function depends on the spatial dimension. This means that, expanding around $\delta=0$, the new modes are those in (\ref{Rmode}) and (\ref{hmode}), plus a term proportional to $\delta$. The contribution coming from the `unperturbed' modes will be of the form
\be
P^{(2)}_0=\int\int dt_1dt_2 \ \II_0=\int \frac{d^{3+\delta} p_1}{\mu^\delta} \int d^{3+\delta} p_2 \ \delta^{(3+\delta)}(\vec{k}+\vec{p}_1+\vec{p}_2) \ f(k,p_1,p_2) \, ,
\ee
where $\mu$ is a physical renormalisation scale.\\
By dimensional analysis we can conclude that
\be
\int \frac{d^{3+\delta} p_1}{\mu^\delta} \int d^{3+\delta} p_2 \ \delta^{(3+\delta)}(\vec{k}+\vec{p}_1+\vec{p}_2) \ f(k,p_1,p_2)=k^{-3}\left(\frac{k}{\mu}\right)^\delta F(\delta) \, ,
\label{dimreg}
\ee
where $F$ is a dimensionless analytic function of $\delta$  with a
pole  of order one in $\delta= 0$
\be
F(\delta)=\frac{F_0}{\delta}+F_1+\mathcal{O}(\delta) \, .
\ee
Notice that we are interested in the late time behaviour, with $k$
superhorizon. In this limit, we suppose that  no residual time
dependence is left in the 2-point function; this is the case when non-adiabatic contribution can
be neglected.
Therefore, expanding around $\delta=0$, the correction can be written
as
\be
P^{(2)}_0(k)=k^{-3}\left[F_0\log\left(\frac{k}{\mu}\right)+\Lambda\right] \, ,
\label{leadorder}
\ee
where $\Lambda$ is a divergent constant. The coefficient $F_0$ can be easily obtained by evaluating the LHS of (\ref{dimreg}) and then by differentiating both the sides of the equation a suitable number of times. However, this is not the end of the story, since there are now the terms proportional to $\delta$ in the corrected wavefunctions; one can verify that they will always be of the form
\be
\delta \, \frac{df_k }{d \delta} = \delta\,\left[\frac{1}{2}\,f_k|_{\delta=0}\,\log(-H\,t)+u(c_s,k,t)\right] \, .
\ee
Let us discuss briefly why contributions coming from the $u(c_s,k,t)$ functions will not provide any logarithmic correction after the integrations are performed. Schematically:
\be
P^{(2)}_u(k)=k^{-3} \ \delta\left(\frac{k}{\mu}\right)^\delta \tilde{F}(\delta) \, .
\ee
Therefore those terms could give contributions only if $\tilde{F}(\delta)$ has a double pole in $\delta$. Still, these corrections to the wavefunctions do not increase the degree of divergence of $\tilde{F}(\delta)$, at least in our cases of interest, so that they will give no contribution to the loop correction.\\
The results are different for the terms proportional to $\log(-Ht)$;
although it could be quite involved, the calculation can be performed
exactly, and the structure results to be
\be
P^{(2)}_\delta(k)=k^{-3} \ \delta\left(\frac{k}{\mu}\right)^\delta
G(\delta) \ \ \ \ \ \ \ \ \ \ \ \ \ \ \ \ \ \ \
G(\delta)=\frac{G_0}{\delta}+G_1+\mathcal{O}(\delta) \, .
\label{g0}
\ee
Therefore the total correction will be given by putting together (\ref{leadorder}) and (\ref{g0})
\be
P^{(2)}(k)=k^{-3}\left[F_0\log\left(\frac{k}{\mu}\right)+G_0+\tilde{\Lambda}\right]
\ee
with $\tilde{\Lambda}$ again a divergent constant. The important
message is that $G_0$ could contribute to the loop correction only if
it has some logarithmic dependence on $k$, otherwise it will be
absorbed into $\tilde{\Lambda}$.

Let us now consider a generic Lagrangian in Fourier space of the form
\be
L=  \plm^2 \, a^{2\nu-1}_{dS}  \left ( f'{}^2 - k^2 \, f^2 \right) +
L_{TSS}\, , \qquad \qquad L_{TSS} = A \, \plm^2 \, a^{2\nu-1}_{dS} \, D(\textbf{k},\textbf{k}',\textbf{k}'')  \, h_k \, f_{k'} \, f_{k''}  \, ;
\ee
where $A$ is a constant and $f$ is some scalar quantity gauge invariant and
 constant at superhorizon scales (e.g. $\mathcal{R}$ for
standard single field inflation, $\zeta$ for solid inflation). The $D$ function encodes all the information about the structure of the spatial derivatives acting on the fields contracted with the polarisation tensor. By solving the linear
equation of motion for $f$, the mode corresponding to the Bunch-Davies
vacuum has the form 
\be
 f_k(t) =\frac{i \pi}{2^\nu \ \Gamma(\nu)}\frac{A_f}{k^{3/2}} \ x^\nu \ H_\nu^{(1)}(x) \, , \qquad \qquad x = - c_s \, k \, t \, ,
\ee
 where $A_f$ is the scale invariant amplitude of the power spectrum of
 $f$. For instance, in single field $\nu=3/2$ while for a solid $\nu=5/2$.
If we use dimensional regularisation, then one can verify that for
each vertex there is a contribution
\be
a^{2\nu-1}(t_{1,2})\left[1-\left(\nu-\frac{1}{2}\right)\delta\log(-Ht_{1,2})\right]
\, ;
\ee
and for each (internal) wavefunction a correction coming from
\be
(-Ht_{1,2})^{\nu+\delta/2}=(-Ht_{1,2})^\nu\left[1+\frac{\delta}{2}\log(-Ht_{1,2})\right]
\, .
\ee
Note that is not necessary to regularize external fields, since every logarithm coming from this functions will be canceled by the corresponding counterterm which renormalises the interaction, see~\cite{Senatore:2009cf}.
If we write the tree level integrand as $\mathcal{I}_0$, then we have to compute the integral of
\be
\II=\II_0\left[1+\delta \ (2-\nu) \ \sum_{i=1}^2\log(-Ht_i)\right] \, .
\ee
The observation made in~\cite{Senatore:2009cf} is that the
contribution coming from the logarithmic corrections of the
wavefunctions is simply a multiplicative factor in front of the tree
level result, namely
\be
\frac{1}{2}\int\int dt_1dt_2 \ \log(H^2t_1t_2) \
\II_0=\log\left(\frac{H}{k}\right)\int\int dt_1dt_2 \ \II_0 \, .
\ee
Following this idea, what we have in the general case is
\be
\delta \ (2-\nu)\int\int dt_1dt_2 \ \log(H^2t_1t_2) \ \II_0=2(2-\nu) \
\delta \ \log\left(\frac{H}{k}\right)\int\int dt_1dt_2 \ \II_0 \, .
\ee
As usual, the tree level contribution is
\be
\int\int dt_1dt_2 \
\II_0=\frac{1}{k^3}\left(\frac{k}{\mu}\right)^\delta F(\delta) \, ;
\ee
thus, the total result is given by
\be
P^{(2)}= \frac{F_0}{k^3}\cdot\left[\log\left(\frac{k}{\mu}\right)+2(2-\nu)\log\left(\frac{H}{k}\right)\right] \, .
\ee
 As a result,  in the case of single-field inflation or when a scalar spectator field
 is present $\nu=3/2$ and no running in $k$ is present. On the
 contrary, for a solid where $\nu=5/2$, or more in general when $\nu\neq 3/2$ the cancellation does not occur anymore.

\bibliographystyle{unsrt}  
\bibliography{biblio}

\end{document}